\documentstyle[aps]{revtex}
\newcommand{\rinv}{\stackrel{\rightarrow}{G_{(0)}^{-1}}}
\newcommand{\linv}{\stackrel{\leftarrow}{G_{(0)}^{-1}}}
\begin{document}
\draft
\title{Quantum Field Kinetics}
\author{ A.  Makhlin}
\address{Department of Physics and Astronomy, Wayne State University,
Detroit, MI 48202}
\date{\today}
\maketitle
\begin{abstract}
Using the general framework of quantum field theory, we derive basic
equations of quantum field kinetics. The main goal of this approach is
to compute the observables associated with a quark-gluon plasma at
different stages of its evolution. We start by rewriting the integral
equations for the field correlators in different forms, depending on
the relevant dynamical features at each different stage.  Next, two
versions of perturbation expansion are considered. The first is best
suited for the calculation of electromagnetic emission from chaotic,
but not equilibrated, strongly interacting matter.  The second version
allows one to derive evolution equations, which are generalizations of
the familiar QCD evolution equations, and provide a basis for the
calculation of the initial quark and gluon distributions after the
first hard interaction of the heavy ions.
\end{abstract}

\noindent {\bf \Large 1. Introduction}

The most ambitious goal of the RHIC and LHC programs is to discover a new
state of the matter -- the quark-gluon plasma (QGP).  Evidence for a QGP
will require a self-consistent analysis of many signals from all stages of
the collision -- initial ``hard'' processes ($\tau < 1/2~{\rm fm/c}$),
the QGP itself, and from the ``cool'' hadronic gas ($\tau \sim 10-30~{\rm
fm/c}$). Thus, a continuous description for all $\tau$ is of more than of
academic interest.  This is a difficult task, and currently, each stage is
described using different approach. Here, we primarily wish to design a
formalism that allows one to describe all stages of the collision,
including the transient ones, using the same technical tools.

The existence of a QGP is inseparable from the process of creation of
the matter it consists of. It is an essentially quantum process .
Therefore an exact definition of the initial state of the system, and
of the observables in the expected final state, is required. Two
difficulties arise: (i) It is unclear how the stable nuclei of the
initial state are build up from quarks and gluons; and (ii) the
expected final state is imperfectly understood.  Thus, the theory
should be adaptable enough to deal with these uncertainties. It must
bridge the logical gap between the language used to describe stable
nuclei and a QGP, as is done phenomenologically in using an
intermediate free parton language.  More specifically, in describing a
transition between the initial state of two stable, well shaped
nuclei, and a system of free quarks and gluons, we encounter one of
the most painful problems of quantum field theory -- initial and final
states of the system can not be described in terms of the same
language.  Before the initial collision, the system is confined and
the vacuum is dominated by quark and gluon condensates, while
afterwards all the condensates are destroyed and the quark-gluon
dynamics is calculated with respect to the perturbative vacuum. This
vacuum itself is a product of the collision.

We may begin the study with the assumption that a high multiplicity of
quarks and gluons has already been created. However, this ``plasma''
may remain out of thermal and chemical equilibrium for a long time,
and may not reach it at all \cite{hotglue}.  In this case, the
calculation of even relatively simple signatures such as dilepton and
photon emission is a nontrivial problem.  An overview of the many
publications on this subject reveals that the QGP has generally been
treated as a totally equilibrated system, and that calculations are
essentially based on the detail balance relationships.

Therefore, we  may conclude that any reasonable  theory should rely neither
on detail balance and  thermal equilibrium, nor  even on the existence of the
same ground state for all stages of the QGP evolution. The theory should
explicitly follow the temporal sequence of the stages, and allow for smooth
transitions between them. The prototype  of such a theory was designed by
Keldysh \cite{Keld} for  non-equilibrium  condensed matter systems with the
aim of deriving a quasi-classical kinetic equation.

The version of the Keldysh technique presented below, and named
``Quantum Field Kinetics'' (QFK), takes an intermediate position
between the theory of scattering and quantum field theory of the
many-particle systems. It contains both as its limits. On the one
hand, it allows one to calculate the inclusive cross-sections and
rates of emission since the summation over all unobserved states is
implicit.  Also, it provides the proper balance between ``real
processes'' and ``radiative corrections'' which results in the
cancellation of infrared divergencies, and makes unnecessary to
introduce artificial intermediate cut-offs. On the other hand, QFK
allows one to calculate local observables for many-particle systems.
As a by-product it recovers microcausality, precisely in the form
which was used in axiomatic field theory.

For any particular problem, QFK-based calculations always begin with
the basic definition of observables in terms of their Heisenberg
operators, the Lagrangian of the theory, and the density matrix of the
initial state.  Several examples, motivated by the expected scenario
in heavy ion collisions and possible probes of the QGP, are described
in Section 2. One may express the observables via field correlators
with differently ordered field operators. The correlators are
naturally arrayed in $2\times 2$ matrices, and we follow the original
Keldysh idea of contour ordering. The correlators obey a system of
matrix equations of the Schwinger-Dyson type, which is easily derived
via functional methods. The equations are reviewed in Sec.3.3.

Starting with the initial form of the matrix equations, it is
straightforward to generate a formal perturbation series for the {\it
probabilities} of inclusive processes (instead of {\it amplitubes} for
exclusive ones) in powers of the coupling constant. This series
reproduces the Feynman expansion diagram for diagram, except that now
each vertex acquires an additional dichotomic index. The way to sum
this perturbation series is always influenced by the physically
motivated renormalization conditions, which may be different from case
to case.

As a preliminary step for the future rearrangement of the perturbation
series, we begin, also following Keldysh, with a rotation of the matrix
basis. We introduce the retarded and advanced Green functions of QCD along
with additional correlators which carry information about the space of states
and derive the equations they obey. In Section~3.4 we show that the latter
allow for a formal solution which comes to be the first approximation of
the QCD evolution equations, if we renormalize the retarded propagators
according to the requirement of light cone propagation.

It is important to note that instead of using the traditional method
of the operator product expansion, we can derive the evolution
equations for the DIS structure functions using the language of the
QFK. This language avoids the parton phenomenology which is needed
when DIS structure functions are used for the computation of the quark
and gluon distributions in first hard collision of the two nuclei.
Thus, QFK provides a firmer footing for the description of heavy ion
collisions.

In Section~4 we discuss different versions of the perturbation expansion.
We show that the ordinary perturbation series in powers of the coupling
constant
is adequate for calculations of systems with slowly varying
macroscopic parameters. We describe the scheme using the example of
dilepton emission from the QGP.  An expansion which is suitable
for the violent impact of
relativistic composite systems requires another kind of expansion, one which
preserves the leading light cone singularity of the propagators.

 \renewcommand{\theequation}{2.\arabic{equation}}
\setcounter{equation}{0}
\bigskip
\noindent {\bf \Large 2. Signals from the quark-gluon plasma}
\bigskip

In this section we begin the design of a technique which will allow us
to calculate different observables associated with QGP.  The latter
should be defined unambiguously at both the theoretical and apparatus
level.  We shall try to trace their origin using only the basic
principles of quantum mechanics for as long as is possible. If
successful, we may then express the signals in terms of the parameters
of the emitting system.  All the systems we shall study are described
by the standard QCD Lagrangian. In order to deal with electromagnetic
probes we include the interaction of charged quarks, $q(x)$, and
leptons, $\psi(x)$, with the photon field, $A^{\mu} (x)$, in the total
renormalized interaction Lagrangian:

\begin{eqnarray} {\cal L}_{int}(x) = e \bar{\psi} (x)
\gamma^{\mu}\psi(x)A_{\mu}(x) +  e\sum_{q,i}\bar{q}_{i}(x)
\gamma^{\mu}A_{\mu}(x)q_{i}(x)+ g_{r}\sum_{q,i}\bar{q}_{i}(x)
t^{a}_{ij}\gamma^{\mu} B^a_{\mu}(x)q_{j}(x)+\nonumber\\
+g_{r}f_{abc}\partial^{\mu}B_{a}^{\nu}(x)
B_{\mu}^{b}(x)B_{\nu}^{c}(x)
+ (g_{r}^{2}/4)f_{abc}f_{agh}B_{b}^{\mu}(x)
B_{c}^{\nu}(x)B_{\mu}^{g}(x) B_{\nu}^{h}(x),
\label{eq:Q2.1}
\end{eqnarray}
where $B_{a}^{\mu}(x)$ is the gluon field, and $ g_{r}$ is a
renormalized strong coupling constant. Other notation is commonly used,
and requires no comment.  The standard counterterms which will be used for
renormalization are as follows:
\begin{eqnarray}
{\cal L}_{CT}(x) =
{Z_3-1 \over2} B^{\mu}_{a}(g_{\mu\nu}\partial^2 - \partial_{\mu}\partial_{\nu}+
\lambda n_\mu n_\nu)B^{\nu}_{a}
-(Z_{1}-1) g_{r}f_{abc}\partial^{\mu}B_{a}^{\nu}(x)
B_{\mu}^{b}(x)B_{\nu}^{c}(x)-\nonumber  \\
-(Z_{4}-1) {g_{r}^{2}\over
4}f_{abc}f_{agh}B_{b}^{\mu}(x) B_{c}^{\nu}(x)B_{\mu}^{g}(x)
B_{\nu}^{h}(x)+(Z_2-1) \bar{q}_{i}(x)i \gamma^{\mu} \partial_{\mu} q_{i}(x)
+(Z_{1F}-1) g_{r}\bar{q}_{i}(x)t^{a}_{ij}\gamma^{\mu}
B^a_{\mu}(x)q_{j}(x)
\label{eq:Q2.2}
\end{eqnarray}
Here, all $(Z-1)$'s are considered as the small parameters. In what follows
we shall either use covariant gauges in the lowest order calculations
when the ghosts  do not contribute yet, or shall work in the axial gauge,
$B^\mu n_\mu=0$,  where they are absent.

The Lagrangian (\ref{eq:Q2.1})- (\ref{eq:Q2.2}) gives rise to the ordinary
S-matrix in the $in$-interaction picture,
\begin{eqnarray}
S = T \exp\{i\int d^4x{\cal L}_{int}(x)\}
\label{eq:Q2.3}
\end{eqnarray}
which is considered to be a limit of the evolution operator that
governs the dynamics of Heisenberg observables.

The initial state for any system can be described by
the density matrix  $\rho_{QCD}$. It is formed due to strong
interactions only, and we shall specify it later.
When the  inclusive cross-sections of photon and dilepton
emission are  chosen as the observables, this density operator should
be augmented by the
projector on the initial vacuum state of photons and leptons,
\begin{eqnarray}
  \rho = \rho_{QCD}\otimes|0_{e\gamma}\rangle\langle0_{e\gamma}|,
\label{eq:Q2.4}
\end{eqnarray}
and we assume that the QGP remains transparent  for photons and leptons
throughout its history. The amplitudes of the transition
from one initial state, $|in\rangle$,
to a final one containing photon or dilepton read as
\begin{eqnarray}
 \langle X|c({\bf k},\lambda)S|in\rangle   \;\;\;   {\rm or} \;\;\;
 \langle X|b(2)a(1)S|in\rangle~ ~ ~,
\label{eq:Q2.5}
\end{eqnarray}
where $c({\bf k},\lambda)$ and $a(J)=a({\bf k}_{J},\sigma_{J})$ and $b(J)$
are the photon, electron and positron annihilation operators. Summing the
squared moduli of these amplitudes over a complete set of uncontrolled states
$|X\rangle$,
and averaging over the initial ensemble, we find the inclusive spectra of
photons and dileptons,
\begin{eqnarray}
{  {dN_{\gamma}} \over {d{\bf k}} }
 = \sum_{\lambda} {\rm Sp} \rho_{in}
          S^{\dag}c^{\dag}({\bf k},\lambda)c({\bf k},\lambda)S,
\label{eq:Q2.6}
\end{eqnarray}
\begin{eqnarray}
{   {dN_{e^{+}e^{-}}}
                \over {d{\bf k}_{1}d{\bf k}_{2}}  }
 =   \sum_{\sigma_{1},\sigma_{2}} {\rm Sp}  \rho_{in}
    S^{\dag}a^{\dag}(1)b^{\dag}(2)b(2)a(1)S.
\label{eq:Q2.7}
\end{eqnarray}
It is easy now to commute the Fock operators with $S$ and $S^{\dag}$.
Due to the QED vacuum projector in the density matrix (\ref{eq:Q2.4}),
only the commutators survive and  Eq.(\ref{eq:Q2.6}) takes the  form:
\begin{eqnarray}
   k^{0}{{dN_{\gamma}}\over{d{\bf k}d^{4}x}} =
   {{ig_{\mu\nu}}\over{2(2\pi)^{3}}} \pi^{\mu\nu}_{10}(-k),
\label{eq:Q2.8}
\end{eqnarray}
where $k^0 >0$, and the polarization tensor
\begin{eqnarray}
 \pi^{\mu\nu}_{10}(-k)=-i\int d^{4}(x-y)e^{-ik(x-y)}
\langle{{\delta S^{\dag}}\over{\delta A^{\mu}(x)}}
        {{\delta S}\over{\delta A^{\nu}(y)}}\rangle~,
\label{eq:Q2.9}
\end{eqnarray}
is the Fourier transform of the  product of
two Heisenberg electromagnetic currents averaged with the density
matrix $\rho_{QCD}$:
\begin{eqnarray}
 \pi^{\mu\nu}_{10}(x,y) =i\langle{ \delta S^{\dag} \over
   \delta A^{\mu}(x) }
        {{\delta S}\over{\delta A^{\nu}(y)}}\rangle=
        i\langle {\bf j}^{\mu}(x) {\bf j}^{\nu}(y)\rangle~ ~ ~.
\label{eq:Q2.10}
\end{eqnarray}
The Heisenberg current (as any other local Heisenberg operator)
can be written down in the two equivalent ways,
\begin{eqnarray}
{\bf j}^{\mu}(x)= S^{\dag}T(j^{\mu}(x)S)\equiv
T^{\dag}(j^{\mu}(x)S^{\dag})S,
\label{eq:Q2.11}
\end{eqnarray}
where
{}~$j^{\mu}(x)=(1/2)\sum_{i,q} e_{q}[\bar{q}_{i}(x)\gamma^{\mu},q_i (x)]$~,
is the  operator of the electromagnetic  current in the
{\it in}-interaction picture.
Relations such as (\ref{eq:Q2.11}) are extremely important
as they allow one to keep
the initial order of the operators through all stages of calculations.
This order is strictly prescribed by the definition of the observables and
may not be changed safely, except under very special circumstances.

Now the dilepton rate of emission (\ref{eq:Q2.7}) takes the form
\begin{eqnarray}
 k_{1}^{0} k_{2}^{0} {{dN_{e^{+}e^{-}}}\over{d{\bf k}_{1}
{d\bf k}_{2}d^{4}x}}=
 -ie^{2}{{L_{\mu\nu}(k_{1},k_{2})}\over{4(2\pi)^{6}}}
{\bf \Delta}^{\mu\nu}_{10}(-k),
 \label{eq:Q2.12}
 \end{eqnarray}
where
$k=k_{1}+k_{2}$, and $L^{\mu\nu}= k_{1}^{\mu} k_{2}^{\nu}+
k_{2}^{\mu} k_{1}^{\nu}-g^{\mu\nu}(k_{1}k_{2}-m_{e}^{2}) $
is the polarization sum  of the lepton spinors. The electromagnetic
correlator
\begin{eqnarray}
   {\bf \Delta}^{\mu\nu}_{10}(-k)=-i\int d^{4}(x-y)
  \langle T^{\dag}(A^{\mu}(x)S^{\dag}) T(A^{\mu}(y)S)\rangle
   e^{-ik(x-y)},
\label{eq:Q2.13}
\end{eqnarray}
is a kind of photon Wightman function averaged over $\rho^{QCD}$. The operator
$A(x)$ of the $in$-interaction picture and the Heisenberg operator ${\bf A}(x)$
are connected via relations similar to (\ref{eq:Q2.11}). In the absence of
radiative corrections to the photon propagation, dynamical equations which will
be derived in the next Section will allow us to rewrite the photon correlator
in Eq.~(\ref{eq:Q2.12}) as following,
\begin{eqnarray}
   {\bf \Delta}^{\mu\nu}_{10}(k)=
 -\Delta_{ret}(k)\pi^{\mu\nu}_{10}(k)\Delta_{adv}(k)
=- \pi^{\mu\nu}_{10}(k)/[k^2]^2 ~~~.
\label{eq:Q2.14}
\end{eqnarray}
In order to derive Eqs.~(\ref{eq:Q2.8}) and (\ref{eq:Q2.12}) we have
assumed that an
explicit separation of long-range and short-range scales  is possible,
and have introduced the emission rates per unit volume,  instead of
the inclusive cross-sections.

The way we proceeded above demonstrates the very simple and general
principle of formulation of the problem: Observables like cross-sections
or the rates of emission  must be expressed in terms of specifically
ordered products of the Heisenberg operators.  If we change the photon and
lepton creation and annihilation operators in the
Eq.~(\ref{eq:Q2.5})--(\ref{eq:Q2.7}) for those of quarks or gluons, and
replace the $ \rho_{in}$ of the QGP by a density matrix for the two
colliding nuclei, we  obtain the starting point for a computation of
the quark and gluon distributions  after the  first hard collision.

We now require a formalism which will allow us to calculate these
quantities. We are to keep in mind that all the operators are driven
by ${\cal L}_{int} $, and that all the information about the initial
state of the system is hidden in $\rho_{in}$.  Nothing else is needed
to solve the problem.

\renewcommand{\theequation}{3.\arabic{equation}}
\setcounter{equation}{0}
\bigskip
\noindent {\bf \Large 3.
           The equations of relativistic quantum field kinetics} \\

\bigskip
\noindent {\underline{\it 3.1. Basic definitions}}
\bigskip

In this study, the calculation of observables such as the emission
rate is based on work by Keldysh\cite{Keld}.  It incorporates
a specific set of exact (dressed) field correlators. These
correlators are products of Heisenberg operators, averaged with the
density matrix of the initial state.  For the quark field they read
\begin{eqnarray}
{\bf G}_{10}(x,y) & =-i\langle{\bf q}(x)\bar{{\bf q}}(y)\rangle ,\;\;\;\;\;\;\;
{\bf G}_{01}(x,y) & = i\langle\bar{{\bf  q}}(y){\bf q}(x)\rangle,\nonumber \\
{\bf G}_{00}(x,y) & =-i\langle T({\bf q}(x)\bar{{\bf q}}(y))\rangle ,
{\bf G}_{11}(x,y) & =-i\langle T^{\dag}({\bf q}(x)\bar{{\bf q}}(y))\rangle ,
\label{eq:Q3.1}
\end{eqnarray}
where $T$ and $T^{\dag}$  are the symbols of the time and anti-time ordering.
They may be rewritten in a unified form,
\begin{eqnarray}
   {\bf G}_{AB}(x,y)=-i\langle T_{c}({\bf q}(x_{A})
                       \bar{{\bf q}}(y_{B}))\rangle~ ~ ~,
\label{eq:Q3.2}
\end {eqnarray}
in terms of a special ordering $T_{c}$ along a contour $C=C_{0}+C_{1}$, the
doubled time axis, with $T$-ordering on $C_{0}$ and $T^{\dag}$-ordering on
$C_{1}$. The operators labelled by `1' are $T^{\dag}$-ordered, and stand
before the $T$-ordered operators labelled by `0'. Recalling that
\begin{eqnarray}
 {\bf q}(x)= S^{\dag}T(q(x)S)\equiv T^{\dag}(q(x)S^{})S,
\label{eq:Q3.3}
\end{eqnarray}
we may introduce the formal operator $S_{c}=S^{\dag}S$,
and rewrite (\ref{eq:Q3.2}) using the operators of the $in$-interaction
picture
\begin {eqnarray}
{\bf G}_{AB}(x,y)=-i\langle T_{c}(q(x_{A})\bar{q}(y_{B})S_{c})\rangle,
\label{eq:Q3.4}
\end{eqnarray}
where, by the definition, the internal variables of $S$ lie on
$C_{0}$, and those of $S^{\dag}$ on $C_{1}$.

The boson Greenians are built in the same manner, i.e., for gluon field,
$B(x)$, and photon field, $A(x)$, we have
\begin{eqnarray}
{\bf D}_{AB}(x,y)=-i\langle T_{c}(B(x_{A})B(y_{B})S_{c})\rangle,\\
\label{eq:Q3.5}
{\bf \Delta}_{AB}(x,y)=-i\langle T_{c}(A(x_{A})A(y_{B})S_{c})\rangle,
\label{eq:Q3.6}
\end{eqnarray}
where the vector and colour indices have been suppressed.

We do not consider the path $C$ to be extended to complex values of the
time $t$, nor to be closed. Moreover, to some extent we take a step backwards
by restoring many elements of the old-fashioned perturbation theory which
deals with retarded and advanced propagators, and incorporates
microcausality in the sense of Yang-Feldman equations \cite{YF}. The
$C$-contour
technique is only a convenient trick to do this in an economic way.
Similarly,  the $T$-ordered Green function is the
sole carrier of physical information only under very special circumstances.
Indeed, by definition
\begin {eqnarray}
{\bf G}_{00}(x,y)=-i {\rm Tr} [\rho  S^{\dag} T(q(x)\bar{q}(y)S)]~ ~ ~.
\label{eq:Q3.7}
\end{eqnarray}
If the density matrix of the system, ~$\rho$, corresponds
to the exact stationary ground state, then in presence of an
interaction the state-vector can acquire only a phase factor, and we obtain
Feynman's Green function:
\begin {eqnarray}
{\bf G}_{F}(x,y)={{-i{\rm Tr}\rho T(q(x)\bar{q}(y)S)}
\over {\langle S \rangle_0} }
\label{eq:Q3.8}
\end{eqnarray}
As we shall see in a while,  the Green functions  ${\bf G}_{00}(x,y)$ and
${\bf G}_{F}(x,y)$ even obey different integral equations.

\bigskip
\noindent {\underline{\it 3.2. The density matrix of the initial state}}
\bigskip

The choice of the density matrix specifies the physical phenomenon
under consideration.  If the
initial state of the system consists of a only few excitations of the
vacuum, then we have the density matrix of a pure state. In this case
we are dealing with
the well known picture of scattering.  Eventually, this situation
is described by a certain set of vacuum expectation values,
and the corresponding density matrix of the pure ground state is
\begin{eqnarray}
\rho_{QCD}=|0_{QCD}\rangle\langle 0_{QCD}|.
\label{eq:Q3.9}
\end{eqnarray}

The various density matrices which we shall use later for the  computation
of quark and gluon production in the first hard  AA-collision, and for the
rates of the photon and dilepton emission  from the QGP, emerge from the
following scenario for the heavy ion collision: In the initial stages of a
collision at RHIC or LHC  energies ($\tau \sim 0.1 fm/c$), hard
impacts take place.  The initial state that precedes these processes
consists of  two stable,
well shaped nuclei moving along opposite directions of the light cone,
and the density matrix for  each of them is the same as in the deep
inelastic e-p or $\mu$A scattering. At this stage, the most interesting
observables are the  quark and gluon distributions after the
first interaction that destroyed the nucleons. The QGP domain  begins somewhat
later when the distributions are already  chaotized, and may be described by
one-particle distributions. The most general density matrix which
simulates any given  form of a one-particle distribution is of the
following form:
\begin{eqnarray}
\rho = \prod_{N}\prod_{p,j}e^{-f_{j}(N,p)a^{\dag}_{j}(N,p) a_{j}(N,p)},
\label{eq:Q3.10}
\end{eqnarray}
where ~$N$~ labels the space cells at the hypersurface of the initial data,
and $n_{j}(N,p)=a^{\dag}_{j}(N,p) a_{j}(N,p) $ is an operator of the number
of partons of type $j$ and quantum number $p$ in the $N$-th cell.
Thus, we completely neglect all correlation effects in the initial phase
space. Introduction of the cells is necessary only for extended objects
without long-scale quantum coherence. For a single hadron they are not
needed.

The density matrix of the Gibbs ensemble of noninteracting quarks and gluons
against a hydrodynamic background is of the same kind as (\ref{eq:Q3.10}),
and obeys the additional condition of being an
entropy extremum. Introducing the
local 4-velocity of continuous media $u^{\mu}(x)$ we may write
\begin{eqnarray}
    \rho_{QCD}=\prod_{N}  \frac{\exp [(-P_{N}u_{N}+\mu_{N}Q_{N})
        /T_{N}]}{Sp[\exp [(-P_{N}u_{N}+\mu_{N}Q_{N})/T_{N}]]}~ ~ ~,
\label{eq:Q3.11}
\end{eqnarray}
where $P^{\mu}_{N}$ and $Q_{N}$ are the total 4-momentum and
(baryonic) charge of free quarks and gluons at temperature $T_{N}$,
and $\mu_{N}$ is the chemical potential in a small 3-volume $V_{N}$ on
the hypersurface of the initial data.

The explicit form of the ``bare'' Greenians which are the basis of
the perturbation  theory is quite evident: The ``vacuum'' Green functions
are  of the standard form,
\begin{eqnarray}
 G^{(0)}_{10,01}(s) =&
 -2 \pi i (\not s+m) \theta (\pm s_{0})\delta (s^{2}-m^{2}),\;\;\;\;\;\;
 G^{(0)}_{00,11}(s)  =& \pm {\not s+m \over s^{2}-m^{2} \pm i0}~, \nonumber \\
D^{(0)\mu \nu}_{10,01}(s)  =& -2 \pi i d^{\mu\nu}(s)
\theta (\pm s_{0})\delta (s^{2})~,
   D^{(0)\mu \nu}_{00,11}(s)   =& \pm {d^{\mu\nu}(s) \over s^{2}\pm i0}~.
\label{eq:Q3.12}
\end{eqnarray}
where the projector $d^{\mu\nu}(s)$  depends upon the choice of gauge.
Usually, the bare Greenians of the ensemble are of the form
\begin{eqnarray}
  G_{AB}(s) = G^{(0)}_{AB}(s)+G_{\beta}(s), \;\;\;
   D^{\mu\nu}_{AB}(s) = D^{(0)\mu\nu}_{AB}(s)+D^{\mu\nu}_{\beta}(s),
\label{eq:Q3.13}
\end{eqnarray}
where the additional terms originating from the $\rho_{in}$,
\begin{eqnarray}
  G_{\beta}(s)= 2 \pi i (\not s +m)\delta(s^{2}-m^{2})
  [\theta (s_{0})n^{(+)}(s)+\theta (-s_{0})n^{(-)}(s)],  \nonumber   \\
     D^{\mu\nu}_{\beta}(s)  = -2 \pi i d^{\mu\nu}(s) \delta(s^{2})
     [\theta (s_{0})f^{(+)}(s)+\theta (-s_{0})f^{(-)}(s)],
\label{eq:Q3.14}
 \end{eqnarray}
manifestly contain the Fermi- or Bose-occupation numbers, $n^{(\pm)}$
or $f^{(\pm)}$, respectively.  They are the diagonal elements of the
density matrix.  All theorems of the Wick--type, which are necessary
for the calculations, can be proven easily for density matrices
like ~(\ref{eq:Q3.10}) and (\ref{eq:Q3.11}).

\bigskip
\noindent {\underline{\it 3.3. The Schwinger-Dyson equations for QFK}}
\bigskip

Except for the matrix form, the Schwinger-Dyson equations for the Heisenberg
correlators remain the same as in any other technique. An elegant and
universal way to  derive them which does not rely on the initial diagram
expansion can be found in Ref.\cite{Bogol}. For the quark field these
equations are of the form
\begin{eqnarray}
{\bf G}_{AB} =
G_{AB}+ \sum_{RS} G_{AR}\circ \Sigma_{RS}\circ {\bf G}_{SB},
\label{eq:Q3.15} \end{eqnarray}
Here the dot stands for convolution in coordinate space, and for
the usual product in  momentum space (providing the system can be  treated
as homogeneous in space and time).   Indeed, the only tool used to derive
these equations was the Wick theorem for  the ordered products of the
operators.  The type of ordering is inessential \cite{Bogol}. Explicit
expressions for  the self-energies are obtained automatically in course of
the derivation of  the  Schwinger-Dyson equations. The quark self-energy
matrix reads
\begin{eqnarray}
  \Sigma_{AB}(x,y)=i(-1)^{A+B}g_{r}^{2}\sum_{R,S=0}^{1} (-1)^{R+S}
   \int d \xi d \eta t^{a} \gamma^{\mu} {\bf G}_{AR}(x,\xi)
   \Gamma^{d,\lambda}_{RB,S}(\xi,y;\eta)
 {\bf D}^{da}_{SA,\lambda\mu}(\eta,x)~.
\label{eq:Q3.16}
\end{eqnarray}
The form the strong $qqB$-vertex that appears is
\begin{eqnarray}
 \Gamma^{d,\lambda}_{SQ,P}(x,y;z)=(-1)^{P+S+Q}{
 {\delta [{\bf G}^{-1}(x,y)]_{SQ} } \over
 {g_{0} \delta {\cal B}^{d}_{\lambda}(z_{P}) }  }~ ~ ~ ,
\label{eq:Q3.17}
\end{eqnarray}
which is the functional derivative of the inverse Greenian of the
quark field with respect to the ``external'' gluon field ${\cal B}(x)$.

The matrix of the photon correlators obeys similar equations,
\begin{eqnarray}
  {\bf \Delta}_{AB} = \Delta_{AB}+ \sum_{RS}
 \Delta_{AR}\circ \Pi_{RS}\circ {\bf \Delta}_{SB},
\label{eq:Q3.18}
\end{eqnarray}
where the electromagnetic polarization operator is
\begin{eqnarray}
{\cal P}^{\mu\nu}_{AB}(x,y)=-i(-1)^{A+B}g_{r}^{2} \sum_{R,S=0}^{1}(-1)^{R+S}
\! \int \! d \xi d \eta \gamma^{\mu}{\bf G}_{AR}(x,\xi)
 E^{\nu}_{RS,B}(\xi,\eta;y) {\bf G}_{SA}(\eta,x),
\label{eq:Q3.19}
\end{eqnarray}
and where the electromagnetic vertex
\begin{eqnarray}
    E^{\lambda}_{RS,P}(x,y;z)=(-1)^{R+S+P}{
 {\delta [{\bf G}^{-1}(x,y)]_{RS} } \over
  {e \delta {\cal A}_{\lambda}(z_{P}) }  }
\label{eq:Q3.20}
\end{eqnarray}
is dressed by the strong interaction. The latter in  turn obeys the equation
\begin{eqnarray}
  E^{\mu}=\gamma^{\mu}+E^{\mu}\circ {\bf G G}\circ {\bf K},
\label{eq:Q3.21}
\end{eqnarray}
with a four-fermion vertex  ${\bf K}$.

The gluon field correlators also obey similar equations
\begin{eqnarray}
{\bf D}_{AB}=D_{AB}+ \sum_{RS} D_{AR}\circ \Pi_{RS}\circ {\bf D}_{SB},
\label{eq:Q3.22}
\end{eqnarray}
where the gluon self-energy has two terms:
\begin{eqnarray}
 \Pi^{\mu\nu,ab}_{AB}(x,y)=-ig_{r}^{2} \sum_{R,S=0}^{1}(-1)^{A+B+R+S}
[\int \! d \xi d \eta {\rm Tr}\gamma^{\mu}{\bf t}^a{\bf G}_{AR}(x,\xi)
 \Gamma^{\nu,b}_{RS,B}(\xi,\eta;y) {\bf G}_{SA}(\eta,x) - \nonumber \\
-\int \! d \xi d \eta V^{\mu\alpha\lambda}_{acf}(x,\xi\eta')
{\bf D}_{AR}^{cc',\alpha\beta}(\xi,\xi')
 {\bf V}^{\nu\beta\sigma}_{RSB;bc'f'}(\xi',\eta,y)
{\bf D}_{SA}^{f'f,\lambda\sigma}(\eta,\eta')] \hspace*{1cm}.
\label{eq:Q3.23}
\end{eqnarray}
The 3-gluon vertex (in coordinate representation) is defined as
 \begin{eqnarray}
 {\bf V}^{\nu\beta\sigma}_{bcf,RSP}(x,y,z)=(-1)^{R+S+P}{
 {\delta [{\bf D}^{-1}(x,y)]^{bc;\nu\beta}_{RS} } \over
  {g_r \delta {\cal B}^{f}_{\sigma}(z_{P}) }  }~.
\label{eq:Q3.24}
\end{eqnarray}
We have omitted the trivial term with the 4-gluon vertex in
Eq.~(\ref{eq:Q3.23}). The explicit
coordinate expression for the bare  3-gluon vertex is
 \begin{eqnarray}
 V^{\nu\beta\sigma}_{bcf,RSP}(x,y,z)=(-1)^{R+S+P}{
 {\delta [D_{(0)}^{-1}(x,y)]^{bc;\nu\beta}_{RS} } \over
  {g_r \delta {\cal B}^{f}_{\sigma}(z_{P}) }  } ,
\label{eq:Q3.25}
\end{eqnarray}
where the inverse Green function $D_{(0)}^{-1}$ denotes
a linearized differential operator of the wave
equation for the gluon in an external gluon field.
The only importance of 3-gluon vertex in this form is that it displays
its local nature,
$V_{ABC}(x,y,z)\sim \delta_{AB}\delta_{AC} \delta(x-y)\delta(x-z)$.
In  momentum representation it may be written as
\begin{eqnarray}
  V^{\alpha\beta\gamma}_{ABC;abc}(p_1,p_2,p_3)=-ig_r\delta_{AB}\delta_{AC}
f^{abc}[g^{\alpha\beta}(p_1 -p_2)^\gamma +g^{\beta\gamma}(p_2 -p_3)^\alpha
+g^{\gamma\alpha}(p_3-p_1)^\beta ]~ ~ ~.
\label{eq:Q3.26}
\end{eqnarray}

The four types of operator ordering which enter Eqs.~(\ref{eq:Q3.1})
are not linearly independent, {\it i.e.}, there exists a set of
relations between the field correlators, and between the
self-energies:
\begin{eqnarray}
G_{00}+G_{11}=G_{10}+G_{01},\;\;\;\;\;\;
 \Sigma_{00}+\Sigma_{11}=-\Sigma_{10}-\Sigma_{01}, \nonumber \\
D_{00}+D_{11}=D_{10}+D_{01},\;\;\;\;\;\;
\Pi_{00}+\Pi_{11}=-\Pi_{10}-\Pi_{01}~ ~ ~.
\label{eq:Q3.27}
\end{eqnarray}
These indicate that only three elements of the $2 \times 2$ matrices  $G,
\Sigma$, {\it etc.} are independent. To remove the overdetermination let us
introduce new functions
\begin{eqnarray}
 G_{ret}  =G_{00}-G_{01},\;\;\;\;\; G_{adv}  =G_{00}-G_{10},\;\;\;\;\;
 G_{1}  =G_{00}+G_{11};  \nonumber \\
 \Sigma_{ret}  =\Sigma_{00}+\Sigma_{01},\;\;\;\;\;
 \Sigma_{adv}  =\Sigma_{00}+\Sigma_{10}, \;\;\;\;\;
 \Sigma_{1}  =\Sigma_{00}+\Sigma_{11},
\label{eq:Q3.28}
\end{eqnarray}
as well as their analogs for bosonic correlators. One of the possible ways
to exclude the extraneous quantities is to use the following unitary
transformation \cite{Keld},
\begin{eqnarray}
\tilde{G}=R^{-1}GR, \;\;\;\;\; \tilde{\Sigma}=R^{-1}\Sigma R ,\;\;\;\;\;
R={{1}\over {\sqrt{2}}} \left| \begin{array}{rc}
                                            1 & 1 \\
                                           -1 & 1  \end{array} \right|.
\label{eq:Q3.29}
\end{eqnarray}
In this new representation, the matrices of the field correlators and
self-energies have a triangle form,
\begin{eqnarray}
  \tilde{G}= \left| \begin{array}{ll}
                                       0 & G_{adv} \\
                                 G_{ret} & G_{1}     \end{array} \right| ,
\;\;\;\;\;  \tilde{M}= \left| \begin{array}{ll}
                                 \Sigma_{1}    & \Sigma_{ret}  \\
                                 \Sigma_{adv}  & 0      \end{array} \right|
\label{eq:Q3.30}
\end{eqnarray}
Applying transformation ~(\ref{eq:Q3.29}) to the matrix
Schwinger-Dyson equations  (\ref{eq:Q3.15}) we may rewrite them  in the
following form:
\begin{eqnarray}
 {\bf G}_{ret} = G_{ret}+ G_{ret}\circ \Sigma_{ret}\circ {\bf G}_{ret} ,
\label{eq:Q3.31}\end{eqnarray}
\begin{eqnarray}
 {\bf G}_{adv} = G_{adv}+ G_{adv}\circ \Sigma_{adv}\circ {\bf G}_{adv}  ,
\label{eq:Q3.32}\end{eqnarray}
\begin{eqnarray}
 {\bf G}_{1} = G_{1}+ G_{ret}\circ \Sigma_{ret}\circ {\bf G}_{1}+
G_{1}\circ \Sigma_{adv}\circ {\bf G}_{adv}+
G_{ret}\circ \Sigma_{1}\circ {\bf G}_{adv}
\label{eq:Q3.33}\end{eqnarray}

There are now two ways to proceed with a further study of Eq.~(\ref{eq:Q3.33}).
If, as typical of condensed matter systems,
one has two well
separated  short- and  long-range scales, then it is reasonable to look for
a kind of differential kinetic equation \cite{Keld}.
The time- and space-derivatives
on its l.h.s. correspond to the slow variations of collective parameters,
while the short-distance
dynamics is absorbed into the collision term.  The existence of two separate
scales here is considered part of the external physical input. It
 must be either self-evident, or
be proven by a separate study (confirmed by observations).
In order to derive the quasi-classical kinetic  equation,  one acts
on Eq.~(\ref{eq:Q3.33}) separately from the left and from the right
with the differential operator of the free wave equation,  and takes
the difference of the
two resulting equations. Further, it is useful to
transform ${\bf G}_{1}$ to  Wigner variables, which effectively separate
the short- and long-range scales. This procedure results in an
integro-differential kinetic equation for the density ${\bf G}_{1}$ in
phase space.  It is differential with respect to the long scale,
and integral with respect to the short scale.

In relativistic Boltzmann-type kinetic theories one encounters specific
difficulties: First, the phase-space Wigner distributions are
no longer positive-definite, and thus do not carry
any direct physical information. Second, for phenomena like
heavy ion collisions, the two scale dynamics is not evident {\it a priori}.
Thus, it is safer to accept the integral equations
(\ref{eq:Q3.31})--(\ref{eq:Q3.33}) as they   are;
however,  in this form they are badly suited
for practical calculations. In a short while we shall show that the exact
equations (\ref{eq:Q3.31})--(\ref{eq:Q3.33}) can be
solved (at least formally).

\bigskip
\noindent {\underline{\it 3.4. The formal solution of the integral equations}}
\bigskip

The solution we shall look for now is equivalent to the rearrangement
of the perturbation series for observables. It is useful
when  specific features of the non-equilibrium system must be taken into
account.  First, let us  introduce two new correlators:
\begin{eqnarray}
 G_{0}= G_{ret}-G_{adv}=G_{10}-G_{01} ,
 \label{eq:Q3.34}\end{eqnarray}
which coincide with the anti-commutator of the quark fields and thus
disappear outside the light cone, and
\begin{eqnarray}
 \Sigma_{0}= \Sigma_{ret}-\Sigma_{adv}=-\Sigma_{10}-\Sigma_{01} .
\label{eq:Q3.35}\end{eqnarray}
which is the commutator of two fermion sources and has the same causal
properties as (\ref{eq:Q3.34}).  The integral equation
for $ G_{0}$ may be derived
by taking difference of  Eqs.~(\ref{eq:Q3.31}) and (\ref{eq:Q3.33}):
\begin{eqnarray}
 {\bf G}_{0} = G_{0}+ G_{ret}\circ \Sigma_{ret}\circ {\bf G}_{0}+
G_{0}\circ \Sigma_{adv}\circ {\bf G}_{adv}+
G_{ret}\circ \Sigma_{0}\circ {\bf G}_{adv}.
\label{eq:Q3.36}\end{eqnarray}
The sum and the difference of Eqs.~(\ref{eq:Q3.33}) and (\ref{eq:Q3.36})
give corresponding equations for the off-diagonal correlators
$G_{10}$ and $G_{01}$:
\begin{eqnarray}
 {\bf G}_{{\stackrel{01}{\scriptscriptstyle 10}}}
= G_{{\stackrel{01}{\scriptscriptstyle 10}}}+
 G_{ret}\circ \Sigma_{ret}\circ {\bf G}_{01,10}+
G_{{\stackrel{01}{\scriptscriptstyle 10}}}\circ
\Sigma_{adv}\circ {\bf G}_{adv} -
G_{ret}\circ \Sigma_{{\stackrel{01}{\scriptscriptstyle 10}}}\circ
{\bf G}_{adv}  .
\label{eq:Q3.37}\end{eqnarray}
Since Eq.~(\ref{eq:Q3.31}) for the retarded propagator may be identically
rewritten in the same form,
\begin{eqnarray}
 {\bf G}_{ret} = G_{ret}+ G_{ret}\circ \Sigma_{adv}\circ {\bf G}_{adv}+
G_{ret}\circ \Sigma_{ret}\circ {\bf G}_{ret} -
G_{ret}\circ \Sigma_{adv}\circ {\bf G}_{adv} ,
\label{eq:Q3.38}\end{eqnarray}
we may use Eq.~(\ref{eq:Q3.37}) and derive the corresponding equations for
the $T$- and  $T^{\dag}$-ordered propagators:
\begin{eqnarray}
 {\bf G}_{{\stackrel{00}{\scriptscriptstyle 11}}} =
G_{{\stackrel{00}{\scriptscriptstyle 11}}}+
G_{ret}\circ \Sigma_{ret}\circ
{\bf G}_{{\stackrel{00}{\scriptscriptstyle 11}}}+
G_{{\stackrel{00}{\scriptscriptstyle 11}}}
\circ \Sigma_{adv}\circ {\bf G}_{adv} + G_{ret}
\circ \Sigma_{{\stackrel{11}{\scriptscriptstyle 00}}}\circ {\bf G}_{adv}.
\label{eq:Q3.39}\end{eqnarray}
This chain of routine transformations reduces equations for all
elements of the matrix correlator $G_{AB}$ to a unified form.
On the one hand, this representation shows that linear relations
between the correlators (or, explicitly, different types of orderings)
hold even for the equations that the correlators obey. On the other hand,
this representation of the equations singles out the role of retarded and
advanced propagators over
all other correlators. In order to understand why their
role is special,  let us transform them further, and  begin by
rewriting of Eq.~(\ref{eq:Q3.33}) for the density of states
${\bf G}_{1}$ identically as:
\begin{eqnarray}
 (1- G_{ret}\circ \Sigma_{ret})\circ {\bf G}_{1}=
G_{1}\circ (1+\Sigma_{adv}\circ {\bf G}_{adv}) +
G_{ret}\circ \Sigma_{1}\circ {\bf G}_{adv}~ ~ ~.
\label{eq:Q3.40}\end{eqnarray}
Since ${\bf G}_{ret}\circ \Sigma_{ret}\circ G_{ret}={\bf G}_{ret}-G_{ret}$,
it is easy to show that
\begin{eqnarray}
 (1+ {\bf G}_{ret}\circ \Sigma_{ret}) (1- G_{ret}\circ
\Sigma_{ret})=1~ ~ ~.
\label{eq:Q3.41}
\end{eqnarray}
Further, we have the following two relations
\begin{eqnarray}
 (1+\Sigma_{adv}\circ {\bf G}_{adv}) = \rinv \circ {\bf G}_{adv}~,\;\;\;\;\;
 (1+ {\bf G}_{ret}\circ \Sigma_{ret})=   {\bf G}_{ret}\circ \linv~ ,
\label{eq:Q3.42}
\end{eqnarray}
where
\begin{eqnarray}
\rinv (x) = i \not \partial_{x}-m,\;\;\;\linv (x) = -i \not\partial_{x}-m,
\label{eq:Q3.43}
\end{eqnarray}
are the left and right differential operators of the Dirac equation,
respectively. Now, multiplying Eq.~(\ref{eq:Q3.40})
by $(1+ {\bf G}_{ret}\circ \Sigma_{ret}) $
from the left, we find the final form of the equation we are looking for:
\begin{eqnarray}
 {\bf G}_{1} = {\bf G}_{ret}\circ  \linv \circ G_{1}
 \circ \rinv \circ {\bf G}_{adv}
+{\bf G}_{ret}\circ \Sigma_{1}\circ {\bf G}_{adv}.
\label{eq:Q3.44}
\end{eqnarray}
Repeating these transformations for the other equations, we obtain the
corresponding forms that are most convenient for subsequent analysis:
\begin{eqnarray}
 {\bf G}_{{\stackrel{10}{\scriptscriptstyle 01}}}  =
{\bf G}_{ret}\circ
\linv \circ G_{{\stackrel{10}{\scriptscriptstyle 01}}}
 \circ \rinv \circ {\bf G}_{adv}
-{\bf G}_{ret}\circ \Sigma_{{\stackrel{10}{\scriptscriptstyle 01}}}
\circ {\bf G}_{adv},
\label{eq:Q3.45}
\end{eqnarray}
\begin{eqnarray}
  {\bf G}_{{\stackrel{00}{\scriptscriptstyle 11}}}  =
{\bf G}_{ret}\circ  \linv \circ G_{{\stackrel{00}{\scriptscriptstyle 11}}}
 \circ \rinv \circ {\bf G}_{adv}
+{\bf G}_{ret}\circ \Sigma_{{\stackrel{11}{\scriptscriptstyle 00}}}
\circ {\bf G}_{adv}.
\label{eq:Q3.46}
\end{eqnarray}
While equations ~(\ref{eq:Q3.37}) and (\ref{eq:Q3.39}) are standard
integral equations which have the unknown function on both sides,
after transformation the unknown function appears only on the l.h.s.,
and we may consider Eqs.~(\ref{eq:Q3.44})--(\ref{eq:Q3.46}) as the formal
representation of the required solution.

At this point, the first and the most naive idea is to ignore
the arrows indicating the direction the differential operators act in, and to
rewrite Eqs.~(\ref{eq:Q3.45})--(\ref{eq:Q3.46}) in momentum representation.
Then the first
term in each of Eqs.~(\ref{eq:Q3.45})  will contain the expression
$(p^{2} - m^{2} )\delta(p^{2}-m^{2} )$, which equals zero. It reflects
simple fact that the off-diagonal correlators $G_{10}$ and $G_{01}$
are solutions of the homogeneous Dirac equation. However, this
approach does not appear to be sufficiently consistent:
we would lose the identity between  Eqs.~(\ref{eq:Q3.45})
and  (\ref{eq:Q3.37}), and make it impossible to generate
the standard perturbative expansion in powers of the
coupling constant. Eqs.~(\ref{eq:Q3.46}) will be corrupted also.

A more careful examination of Eqs.~(\ref{eq:Q3.45})-(\ref{eq:Q3.46})
shows that all four dressed correlators ${\bf G}_{AB}$ can be found as
the formal solution of the retarded Cauchy problem, with bare Greenians
as initial data and the self-energies as the sources. Indeed,
integrating the first term of each these equations twice by parts, we find
for all four elements of ${\bf G}_{AB}$
\begin{eqnarray}
{\bf G}_{{\stackrel{10}{\scriptscriptstyle 01}}}(x,y)=
\int d \Sigma_{\mu}^{(\xi)}
 d \Sigma_{\nu}^{(\eta)} {\bf G}_{ret}(x,\xi)\gamma^{\mu}
G_{{\stackrel{10}{\scriptscriptstyle 01}}}(\xi.\eta)\gamma^{\nu}
{\bf G}_{adv}(\eta,y)
-\int d^{4}\xi d^{4} \eta {\bf G}_{ret}(x,\xi)
\Sigma_{{\stackrel{10}{\scriptscriptstyle 01}}}(\xi.\eta)
{\bf G}_{adv}(\eta,y) ,
\label{eq:Q3.47}
\end{eqnarray}
\begin{eqnarray}
{\bf G}_{{\stackrel{00}{\scriptscriptstyle 11}}}(x,y)=
\!\!\int\!\! d\Sigma_{\mu}^{(\xi)}
 d \Sigma_{\nu}^{(\eta)} {\bf G}_{ret}(x,\xi)\gamma^{\mu}
G_{{\stackrel{00}{\scriptscriptstyle 11}}}(\xi,\eta)\gamma^{\nu}
{\bf G}_{adv}(\eta,y)\!
-\! \int \!\! d^{4}\xi d^{4} \eta {\bf G}_{ret}(x,\xi)
[\pm G_{0}^{-1}+ \Sigma_{{\stackrel{11}{\scriptscriptstyle 00}}}(\xi.\eta)]
{\bf G}_{adv}(\eta,y),
\label{eq:Q3.48}
\end{eqnarray}
At first glance it is ambiguous to use the local differential operator
$G_{0}^{-1}$ in Eq.(3.48) without indication of  direction it acts.
Nevertheless,  as follows from (\ref{eq:Q3.42}),
\begin{eqnarray}
{\bf G}_{ret} \circ \rinv \circ {\bf G}_{adv} -{\bf G}_{ret} \circ \linv \circ
{\bf G}_{adv} =
 (1+ {\bf G}_{ret}\circ \Sigma_{ret})\circ {\bf G}_{adv}-
{\bf G}_{ret}\circ (1+\Sigma_{adv}\circ {\bf G}_{adv})= \nonumber\\
=-{\bf G}_{0}+ {\bf G}_{ret}\circ \Sigma_{0}\circ {\bf G}_{adv}=0, \nonumber
\end{eqnarray}
and both directions lead to the same answer.

The corresponding equations for boson fields are obtained by replacing
${\bf G}_{AB} \rightarrow {\bf D}_{AB}$ and ${\bf M}_{AB} \rightarrow
\Pi_{AB}$, and $\gamma^{\mu} d \Sigma_{\mu} \rightarrow
{\stackrel{\leftrightarrow} {\partial}}^{\mu} d \Sigma_{\mu}$ in
Eqs.~(\ref{eq:Q3.34}) through (\ref{eq:Q3.48}).

The equations (\ref{eq:Q3.47}) and (\ref{eq:Q3.48}) as well as
their copies for boson fields are
the basic equations of relativistic quantum field kinetics.  They are
identical to the initial system of Schwinger-Dyson equations, but have
the advantage that the time direction is explicitly singled out.
There are two terms of different origin that contribute to any
correlator (and, consequently, to any observable).
The first term retains some memory about the initial data.
The length of time for which this memory is kept depends on
the retarded and advanced propagators.
The second term describes the current dynamics of the system. A comparison of
these two contribution allows one to judge if the system indeed has two scales.

\renewcommand{\theequation}{4.\arabic{equation}}
\setcounter{equation}{0}
\bigskip
\noindent {\bf \Large 4.~Perturbation theory for quantum field kinetics}\\
\medskip

Any reasonable diagram technique should allow one to assemble certain
subsequences of the bare diagrams into larger pieces, {\it viz.}, the
irreducible elements of the skeleton diagrams, such as exact
Greenians, self-energies and vertices.  The initial standard form of
the Schwinger--Dyson equations (\ref{eq:Q3.15}), (\ref{eq:Q3.16}) and
(\ref{eq:Q3.22}), (\ref{eq:Q3.23}) is an example of such selective
summation.  To derive these equations we have employed the functional
approach, but any dressed field correlator, self-energy or vertex
contains the infinite series of its perturbative expansion in powers
of the coupling constant.  We can reconstruct the formal perturbation
series starting from the skeleton form of the exact Schwinger-Dyson
equations. However, it is not expedient to proceed formally and we
shall not attempt it here. For example, an explicit series of radiative
corrections to the free asymptotic propagation is never retained in
calculations.  In the course of renormalization this series is
absorbed into the physical mass of the field.  The theory of the deep
inelastic e-p scattering that is based on the operator product
expansion (OPE) selectively sums perturbation series in order to
emphasize the dominance of light cone dynamics, and introduces its own
irreducible elements, the structure functions. Just as the masses of
free particles, they are taken from the data, and one should not try
to obtain them directly from a perturbative expansion.  Thus a
physical ({\it versus} mathematical) study begins when we make the
decision on how to sum the perturbation series or, equivalently, what
skeleton elements are not to be expanded in a series. The final choice
is the matter of taste.

We wish now to perform the perturbative expansion of the skeleton
equations (\ref{eq:Q3.15}), (\ref{eq:Q3.18}) and (\ref{eq:Q3.22}).
Their matrix form accounts for a possible instability of the
initial ground state with
respect to interaction, but their mathematical form is precisely the
same as in standard field theory where the initial ensemble
corresponds to a pure and stable ground state.  Any other structure would be
surprising, as only the ordering of the operators, regardless of its
type, is important. The presence of additional indices does not change
the topology of the diagrams.

Eqs.  (\ref{eq:Q3.15}), (\ref{eq:Q3.18}) and (\ref{eq:Q3.22}),
supplemented by the expressions for the self-energies, can be iterated
in a standard way, giving rise to a formal perturbation series in the
powers of coupling constant.  Graphically, the series reproduces the
Feynman diagram expansion with only one difference: each vertex
acquires additional index.  This kind of expansion proves useful when
we solve local problems, such as photon or dilepton emission from a
system with slowly varying macroscopic parameters.  These parameters
are built into the definition of the density matrix and should not
change over the time of emission.

We note, however, that this direct expansion is not most rewarding
when the composite system undergoes a strong and short impact. Its
parameters will change dramatically within a very short time, and the
memory about the density matrix of the true initial state is obviously
lost.  It is then profitable to rearrange Eqs.~(\ref{eq:Q3.18}) and
(\ref{eq:Q3.22}) (for example, ``rotating'' the basis of correlators,
as in Eq.~(\ref{eq:Q3.30})) in order to emphasize other details of the
dynamical process.  Since we are interested in processes on the light
cone, the leading singularity of the retarded propagators is the main
element which should be preserved in course of renormalization.

\bigskip
\noindent
{\underline{\it 4.1. Perturbation theory for the local emission problem}}
\bigskip

The quantity which we shall choose to calculate in this section is the
observable rate of the electromagnetic emission (real $\gamma$ or
virtual $\gamma^*$).  It is expressed via the tensor of the
electromagnetic polarization ${\cal P}^{\mu\nu}_{10}(-k)$ of the QGP,
given by the Eq.~(\ref{eq:Q3.19}).  Let us expand it in a perturbation
series up to order $\alpha_s$.  The polarization tensor ${\cal
P}^{\mu\nu}$ contains three irreducible elements: two quark
correlators and the electromagnetic vertex dressed by the strong
interaction. They are given via their functional definitions, and obey
the inhomogeneous integral equations
(\ref{eq:Q3.15})--(\ref{eq:Q3.21}).  The free terms of these equations
correspond to the bare quark correlators, averaged over the ensemble
represented by the density matrix (of non-interacting particles as a
first approximation).  The density matrix introduces occupation
numbers which define the weights of initial and final states for
elementary processes.

The Born term emerges when all the irreducible elements are
considered as bare:
\begin{equation}
 {\cal P}^{\mu\nu}_{Born}(-k)=ie^{2} N_{c} \int{d^{4}p \over (2\pi)^{4} }
 \gamma^{\mu}G_{10}(p-k)\gamma^{\nu} G_{01}(p).
 \label{eq:Q4.1}
\end{equation}
To obtain corrections of order $\alpha_s$ to the electromagnetic
polarization, one should begin by iterating
Eqs.~(\ref{eq:Q3.15})--(\ref{eq:Q3.21}) to the same order.  This
iteration assumes that we write down equations (\ref{eq:Q3.15}) for
the correlators with the bare Greenians and with self-energies computed
to the first non-vanishing order. The vertex is considered to be bare.
As a result of these approximations, we obtain the first group of
radiative corrections.  A second group appears if we keep field
correlators bare, but include the first order corrections to the
vertex. Eventually, such an expansion results in the following
expression:
 \begin{eqnarray}
 {\cal P} ^{\mu\nu}_{10}(-k)=ie^{2} N_{c} \int{d^{4}p \over (2\pi)^{4} }
 \gamma^{\mu}{\bf G}_{10}(p-k)\gamma^{\nu} {\bf G}_{01}(p)
 +e^{2}g^{2}N_{c}C_{F}\sum_{R,S=0}^{1} (-1)^{R+S} \times \nonumber \\
\times \int{d^{4}pd^{4}q \over
 (2\pi)^{8}} D_{SR}(q)\gamma^{\mu}G_{1R}(p-k)\gamma^{\lambda}
 G_{R0}(p+q-k)\gamma^{\nu}G_{0S}(p+q)\gamma_{\lambda}G_{S1}(p)
 \label{eq:Q4.2}
 \end{eqnarray}
where the correlators ${\bf G}_{10}$ and ${\bf G}_{01}$
should be taken in the form
\begin{equation}
 {\bf G}_{01,10} = G_{01,10}- G_{ret} M_{01,10}  G_{adv} +
  G_{ret} M_{ret}  G_{01,10}+ G_{01,10} M_{adv} G_{adv}~.
 \label{eq:Q4.3}
\end{equation}

All diagrams contributing to ${\cal P}_{10}(-k)$ up to order
$\alpha_s$ are depicted in Fig.\ref{Fig1}.  The sum of these diagrams is the
squared modulus of the coherent sum of amplitudes of the real
processes of photon emission with first virtual corrections. These
amplitudes are given at Fig.\ref{Fig2}.  The first process is the direct
annihilation of a bare $q \bar{q}$ pair to a (virtual) photon, the
next three diagrams are radiative contributions to the same process.
Next, there follow four diagrams of $q \bar{q}$-annihilation
accompanied by gluon emission and absorption, and two diagrams for
Compton scattering of a quark (or antiquark) and a gluon with photon
emission. The first loop and the last four loops in Fig.\ref{Fig1} are due to
real processes, while the others are due to vertex and mass
corrections.  The correlators corresponding to initial or final states
are marked by dashed lines. Lines with arrows correspond to retarded
propagators, and arrows inside the loops show retarded self-energies.

A complete calculation is a subject of separate paper.  Here we shall
only outline the main ideas and results.  The technique of QFK allows
one to perform all calculations without the assumption of thermal
equilibrium.  Not only can the statistical weights of the initial and
final states of the real processes be of arbitrary form (though
compatible with slowly varying macroscopic parameters), but we can
also find all the virtual corrections consistently with the same
quark--gluon background distributions.  The general rule is that
infrared divergences caused by the integration over the low momenta
domain (of the Bloch-Nordsieck type) cancel out between real and
virtual processes, despite the nontrivial population of the phase
space.  It is not even necessary to introduce any artificial
intermediate cut-offs.  Formally, for any given $\alpha_s$-order the
collinear divergences survive, exactly because the phase space of the
initial and final states is populated, and the balance required by the
KLN theorem does not hold. However, the physical origin of these
divergences allows one to conclude that they are shielded by the quark
and gluon distributions: infinitely long free propagation in the
dense system is impossible.

\bigskip
\noindent
{\underline{\it 4.3. Perturbation expansion for light cone processes}}
\bigskip

As was already mentioned in Sections~1 and 2, QFK is equally applicable
to the description of any inclusive processes, including those dominated
by the light cone dynamics. These are deep inelastic e-p scattering,
or deep inelastic collisions of two hadrons or nuclei.
In this case the inclusive cross-section of DIS is expressed via
the same electromagnetic polarization tensor, but the kinematic
region is different.

The cross-section for inclusive quark and gluon production, which
might initiate evolution towards a QGP, is also expressed via the
off-diagonal self-energies of the colliding system. For example, for
gluon production, we have
\begin{equation}
{ dN_g \over d{\bf p}}=\sum_{\lambda,a} Sp \rho_{in}
          S^{\dag}c^{\dag}({\bf p},\lambda,a)c({\bf p},\lambda,a)S
 =\sum_{\lambda,a} \int d^{4}x d^{4}y {e^{-ip(x-y)} \over (2\pi)^3 2p^0}
 \epsilon_\mu^{(\lambda)}\epsilon_\nu^{(\lambda)}[-i\Pi_{aa}^{01,\mu\nu}(x,y)]
\label{eq:Q4.4}
\end{equation}
where the gluon polarization tensor $\Pi_{01}$ is obtained by averaging with
the density matrix of the system of the two colliding nuclei. The latter is the
direct product of two independent density matrices for each nucleus.  Thus,
to first approximation,
\begin{eqnarray}
  \Pi^{\mu\nu}_{01}(p)=ig_{0}^{2}
\{\int  {d^4 kd^4 q \over (2\pi)^4} \delta(k + q - p)
 \{ V^{\mu\rho\sigma}_{acd}(k+q,-q,-k)
{\bf D}_{01,dd'}^{(A)\sigma\beta}(k)
 V^{\nu\lambda\beta}_{bc'd'}(-k-q,q,k)
 {\bf D}_{10, f'f}^{(B)\rho\lambda}(q)]+
(A \leftrightarrow B) \}.
\label{eq:Q4.5}\end{eqnarray}
where each of the gluon field correlators is averaged with the density matrix
of one nucleus, $A$ or $B$.
The pre-collision dynamics of the field correlators
${\bf D}_{01,dd'}^{(A)\sigma\beta}(k)$
and  ${\bf D}_{10,f'f}^{(B)\rho\lambda}(q)$
is naturally described by equations similar to
(\ref{eq:Q3.46}). The conditions for nuclear  propagation along the light
cone allows us to suggest that the retarded propagators
contribute to the process
mainly via their light cone singularity, and can be used without radiative
corrections as a first approximation. In this case we may use the form
\begin{eqnarray}
{\bf D}^{(J)}_{01,10}=D^{(J)\#}_{01,10}+D^{(J)*}_{01,10}-
{\bf D}^{(J)}_{ret}\Pi^{(J)}_{01,10}
{\bf D}^{(J)}_{adv}~ ~ ~ ,
\label{eq:Q4.6}
\end{eqnarray}
where $D^{(J)*}_{01,10}$ describes the population of the gluon states
used as initial data (at, for example, some scale $Q_{0}^{2}$),  and
$D^{(J)\#}_{01,10}$ is the density of gluon states in the initially
unpopulated continuum.

The full description of the problem and the calculations
is the subject of separate paper. Only the main ideas and results
will be outlined here.  For the sake of simplicity, let us consider
the case of pure glue-dynamics. In that case, Eq.~(\ref{eq:Q4.5})
yields the following {\it evolution equation}:
\begin{eqnarray}
  \Pi^{\mu\nu}_{01}(p)=-ig_{r}^{2} \int  {d^4 k \over (2\pi)^4}
  V^{\mu\alpha\nu}_{acf}(p,k-p,k)
\left[D_{ret}(k) \Pi_{10}(k) D_{adv}(k)\right]^{\alpha\beta}
V^{\nu\beta\sigma}_{bc'f'}(-p,p-k,-k)
 D_{10, f'f}^{\#\lambda\sigma}(k-p)
\label{eq:Q4.7}\end{eqnarray}
If this equation is rewritten in light cone variables, and
projected onto the specific
measurement of the unpolarized e-p DIS experiment, it  reproduces
the well known Altarelli-Parisi evolution equation for the gluon structure
function. If $w_1(p)$ is defined as the
scalar coefficient in the decomposition
\begin{equation}
 \Pi^{\mu\nu}(p) =g^{\mu\nu}w_1(p)+{p^\mu p^\nu \over p^2}w_2(p)+...~
{}~ ~,
\label{eq:Q4.8}\end{equation}
then the gluon  structure function of the LLA reads
\begin{equation}
 G(x,Q^2)= G(x,Q_{0}^{2})+{V_{lab}P^+ \over (2\pi)^3}
\int_{Q_{0}^{2}}^{Q^{2}} dp_{t}^{2} \int dp^{-}
{ip^+ w^{01}_{1}(p)\over (p^2)^2}~ ~ ~,
\label{eq:Q4.9}
\end{equation}
where $G(x,Q_{0}^{2})$ is the phenomenologic ``initial'' distribution at
the scale $Q_{0}^{2}$.

Here, the novel point is that equations like (\ref{eq:Q4.7}) are of the
evolution type, even in their initial coordinate form. Further, they
may be derived without reference to any particular process. The
equations are of a ladder form, and the ordering of the ladder cells
by the Feynman $x$ and virtuality $Q^2$ is a consequence of the
initial retarded ordering: lower $x$ and higher $Q^2$ correspond to
later times in the evolution.  A parallel study of the two
processes, p-e DIS, and p-p or A-A collisions, shows that one may
use the new evolution equations to obtain quantities (like $w^{01}_{1}(p)$)
which are common to these two processes. Further, one avoids intermediate
parton language.

\bigskip
\noindent {\bf \Large 7. Conclusions}
\medskip

This paper presents a formalism which allows us to solve various
problems of the QGP dynamics using the same technique.  On the one
hand, it allows one to calculate the rates of such processes as photon
and dilepton emission, or heavy quarks production in the QGP.  On the
other hand, it is a good tool for the study of extremely
non-equilibrium process, such as the first hard collision of two heavy
ions.

{}From the most general point of view, this technique is capable of
replacing the standard Feynman approach, in any situation where the
latter is applicable. In QFK, we begin with the density matrix
generated by the Fock operators from the pure vacuum state of
non-interacting fields. The reader can easily rewrite some chapters
from a standard textbook on QED in terms of this approach as an
exercise. In some points the new approach is physically more intuitive
than the Feynman technique: at the tree level, retarded propagators
appear naturally in every place where we usually put them in by hands
(following common sense).  The cutting rules and unitarity come to be
a trivial consequence of the matrix structure: both $S$ and $S^{\dag}$
contribute to any observable from the very beginning.

For the case of true thermal equilibrium, this kinetic approach also
reproduces all that is obtainable from the standard Matsubara
formalism. All global relations like the FDT come to be a trivial
consequence of thermal equilibrium and the matrix structure of the
Schwinger-Dyson equations, but none of global relations are is used in
the definition of the observables.  For the Gibbs ensemble, QFK
describes dynamical fluctuations.

The essential new point is that the theory is not confined to a specific form
of the one-particle distribution. The plasma may be far out of thermal
and/or chemical
equilibrium. Nevertheless we can consider both real and virtual
processes  in a  self-consistent manner.  Usually this consistency
is lost in computer
simulations of the emission from the non-equilibrium plasma \cite{GL}:
only
tree level matrix elements are taken into account. In the next paper we shall
give an example:  for dilepton emission from a non-equilibrium ``plasma,'' a
self-consistent account of the radiative corrections changes the answers from
what is obtained using a naive cut-off of the infrared
singularities.  We have also found a significant difference in
the case of an equilibrated
plasma.

The QFK approach allows one to formulate new principles for computing
the distributions of quarks and gluons created in the first hard
interaction of two heavy ions at high energies. It essentially employs
an initial resummation of the perturbation series for the
probabilities, and allows one to describe two different high energy
processes, {\it viz.}, e-p scattering and nuclear interactions, in the
same terms. These processes are considered as two versions of the same
phenomenon -- deeply inelastic scattering of composite systems. The
calculations can be performed without reference to parton
phenomenology.

\noindent {\bf Acknowledgements}

I am indebted to G. Brown and the Nuclear theory group at
SUNY at Stony Brook for support during the spring and summer of 1994
when this work was began.
I am grateful to L. McLerran, A.H. Mueller, E.Shuryak, J.Smith, A.Vainshtein
R.Venugopalan and G. Welke for many stimulating discussions.

A part of this work
was supported by the U.S. Department of Energy under Contract
No. DE--FG02--94ER40831.


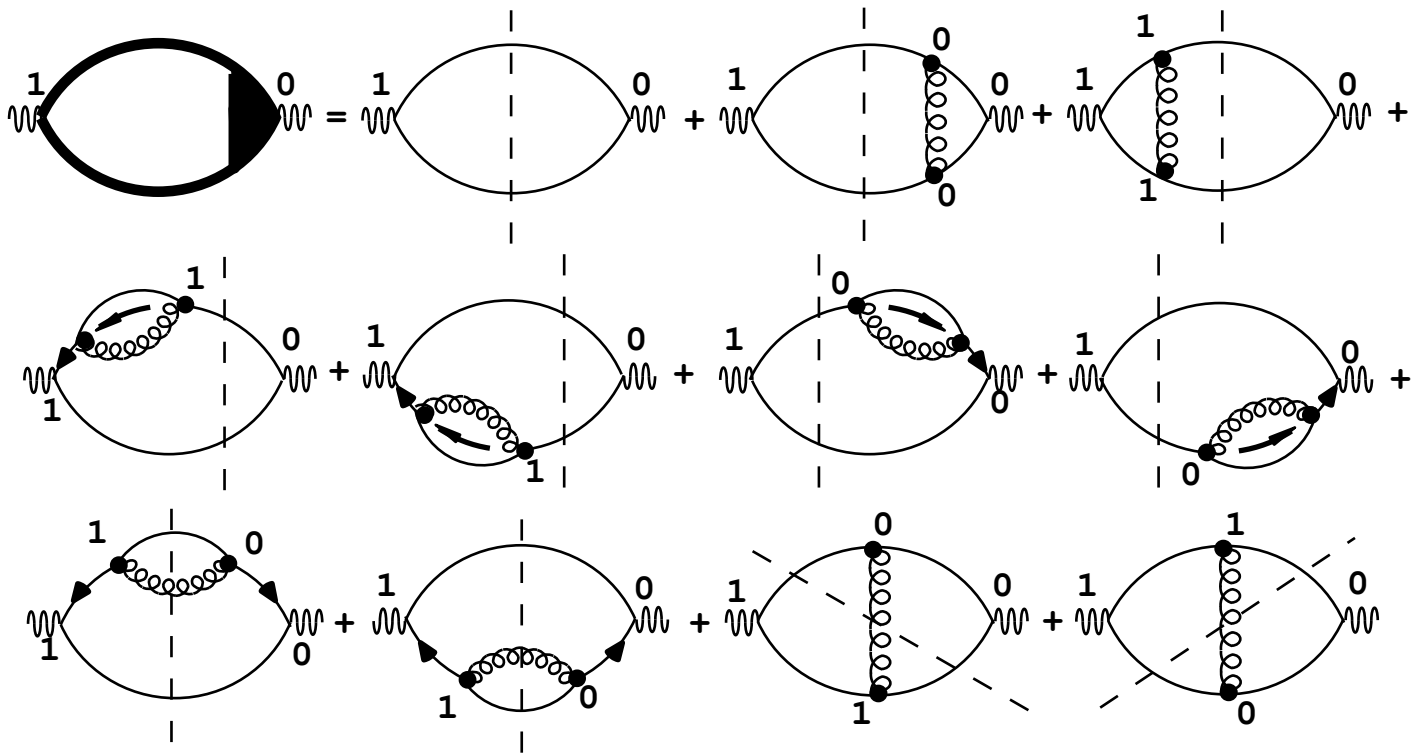
 \begin{figure}
 \caption{Diagrams for the rate of dilepton production.
The dashed line crosses the Greenians representing densities of the
initial or final states. Arrows label
the retarded and advanced propagators and show the latest time.}
 \label{Fig1}
 \end{figure}
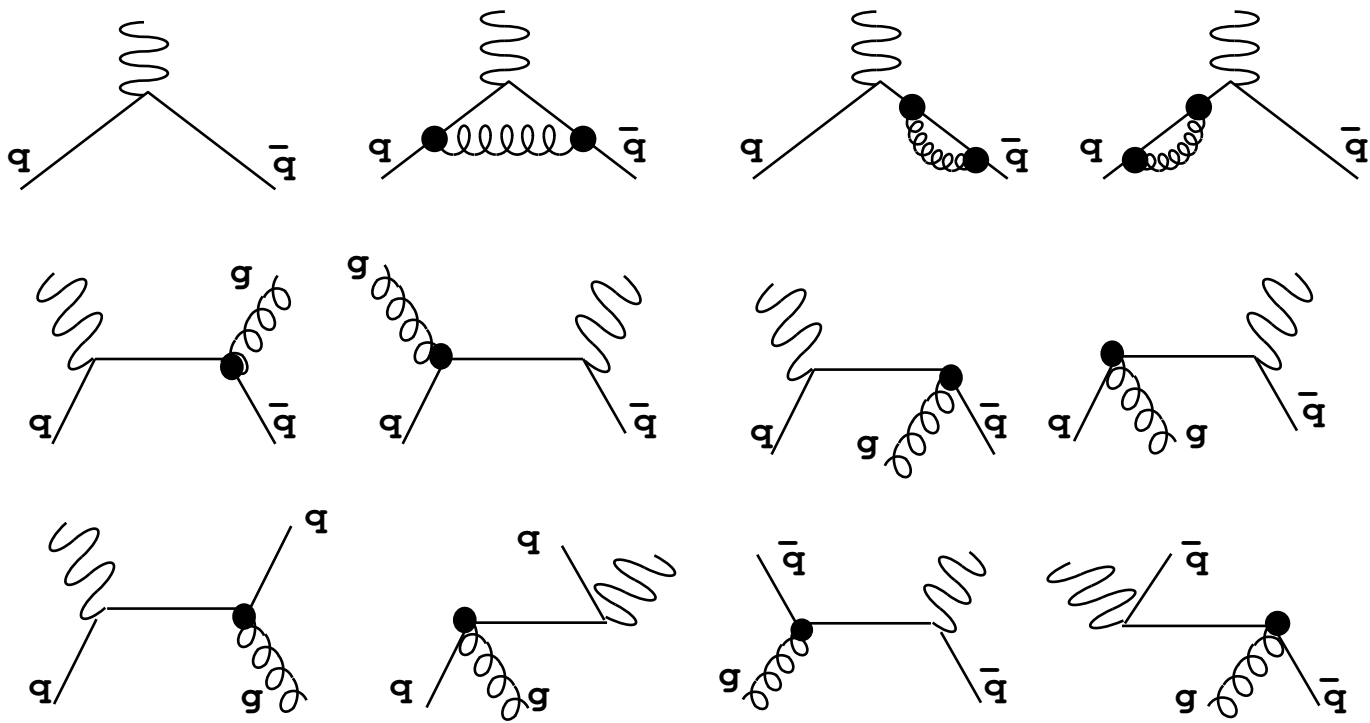
\begin{figure}
 \caption{Processes participating in the dilepton production
in the $\alpha_s$-order.  Notation the same as in Fig.1.}
 \label{Fig2}
 \end{figure}

\end{document}